\documentclass[useAMS,usenatbib,usegraphicx]{mn2e}

\usepackage{journals}
\usepackage[symbol]{footmisc}
\usepackage{url}
\usepackage[usenames]{color}
\usepackage{graphicx}
\usepackage{multirow}

\definecolor{grey}{rgb}{0.5,0.6,0.7}

\title[A universal galaxy photometric relation]{A universal ultraviolet-optical colour--colour--magnitude 
relation of galaxies\thanks{The best-fitting photometric
relation coefficients and other supporting technical information are available 
at the project web-site: http://specphot.sai.msu.ru/galaxies/}}
\author[Chilingarian \& Zolotukhin]{Igor V. Chilingarian$^{1,2}$\thanks{E-mail: chil@sai.msu.ru
(IC); iz@sai.msu.ru (IZ)}, and Ivan Yu. Zolotukhin$^{3,4,2}$\\
$^{1}$Centre de Donn\'ees astronomiques de Strasbourg, Observatoire
astronomique de Strasbourg, Universit\'e de Strasbourg,\\ CNRS UMR~7550, 11
rue de l'Universit\'e, 67000 Strasbourg, France\\
$^{2}$Sternberg Astronomical Institute, Moscow State University, 13 Universitetsky prospect, Moscow, 119992, Russia\\
$^{3}$Observatoire de Paris, LERMA, UMR~8112, 61 Av. de l'Observatoire, 75014 Paris,
France\\
$^{4}$Observatoire de Paris, VO Paris Data Centre, 61 Av. de l'Observatoire, 75014 Paris,
France\\
}
\begin{document}

\date{Accepted 2011 Sep 15. Received 2011 Sep 15; in original form 2011 Feb 6}

\pagerange{\pageref{firstpage}--\pageref{lastpage}} \pubyear{2010}

\maketitle

\label{firstpage}

\begin{abstract} 
The bimodal galaxy distribution in the optical colour--magnitude diagram
(CMD) comprises a narrow ``red sequence'' populated mostly by early-type
galaxies and a broad ``blue cloud'' dominated by star-forming systems.
Although the optical CMD allows one to select red sequence objects, neither
can it be used for galaxy classification without additional observational
data such as spectra or high-resolution images, nor to identify blue
galaxies at unknown redshifts. We show that adding the near ultraviolet
colour (\emph{GALEX NUV} $\lambda_{\mbox{eff}}=227$~nm) to the optical
($g-r$ vs $M_r$) CMD reveals a tight relation in the three-dimensional
colour--colour--magnitude space smoothly continuing from the ``blue cloud''
to the ``red sequence''. We found that 98~per~cent of 225\,000 low-redshift
($Z<0.27$) galaxies follow a smooth surface $g-r = F(M_r, NUV-r)$ with a
standard deviation of 0.03--0.07~mag making it the tightest known galaxy
photometric relation given the $\sim0.9$~mag range of $k$-corrected $g-r$
colours. Similar relations exist in other NUV--optical colours. There is a
strong correlation between morphological types and integrated $NUV-r$ 
colours of galaxies, while the connection with $g-r$ is ambiguous. Rare galaxy
classes such as E+A or tidally stripped systems become outliers that occupy
distinct regions in the 3D parameter space. Using stellar population models
for galaxies with different star formation histories, we show that (a) the
($NUV-r, g-r$) distribution at a given luminosity is formed by objects
having constant and exponentially declining star formation rates with
different characteristic timescales with the red sequence part consistent
also with simple stellar population; (b) colour evolution for exponentially
declining models goes along the relation suggesting a weak evolution of
its shape up-to a redshift of 0.9; (c) galaxies with truncated star
formation histories have very short transition phase offset from the
relation thus explaining the rareness of E+A galaxies. This relation can be
used as a powerful galaxy classification tool when morphology remains
unresolved. Its mathematical consequence is the possibility of precise and
simple redshift estimates from only three broad-band photometric
points. We show that this simple approach being applied to \emph{SDSS} and
\emph{GALEX} data works better than most existing photometric redshift
techniques applied to multi-colour datasets. Therefore, the relation can 
be used as an efficient search technique for galaxies at intermediate 
redshifts ($0.3<Z<0.8$) using optical imaging surveys.
\end{abstract}


\section{Introduction}

Understanding observational aspects of galaxy evolution requires
to classify them in regard to various properties such as morphology,
luminosity, stellar population characteristics, internal dynamics. In the
present-day era of deep wide-field imaging surveys, there is a need in
efficient mechanisms of galaxy classification and selection using minimal
available information. Colour--colour and colour--magnitude diagrams (CMD)
have been traditionally used for this purpose.

In the optical CMD ($g-r, M_r$) \citep{Strateva+01,BGB04,Blanton+03b}, the
very narrow ``red sequence'' \citep{VS77} ($\sigma (g-r) \approx 0.04$~mag)
formed mostly by elliptical and lenticular galaxies is used to identify
early-type members of galaxy clusters because at low redshift it moves as a
whole remaining tight in the colour space. However, optical CMDs cannot be
used for detailed classification of galaxies, either for the selection of
other than red galaxies because of several degeneracies: (1) there is no
unambiguous connection between the galaxy morphological type and its
position on the CMD; (2) the red part of the CMD is contaminated by
$\sim25$~per~cent\ with late-type galaxies having weak ongoing star
formation (SF) attenuated by the dust; (3) the blue cloud overlaps with the
loci of ``E+A'' poststarburst galaxies \citep{DG83} (PSG) having blue
colours, early-type morphology, often disky kinematics \citep{CDRB09} but no
or weak ongoing SF.

In the $(NUV-r, M_r)$ space, both the red sequence and the blue cloud become
pronounced but quite broad ($\sigma(NUV-r)\sim2$~mag) sequences
\citep{Wyder+07}. Such a width of the red sequence is due to the UV flux
sensitivity to even small fractions of young stars that was shown to be
connected to the environment of early-type galaxies \citep{Kaviraj+07}. At
the same time, (1) the sequences are too broad to use them for the efficient
photometric selection of galaxies; (2) there is still an ambiguity between
the $NUV-r$ colour and a galaxy morphological class as well as the presence
of ongoing SF: PSGs still reside in the blue cloud.

\section{The UV--optical galaxy photometric sample}

\subsection{Catalogue construction}

Using Virtual Observatory data mining, we constructed a photometric sample
of $\sim$225\,000 galaxies excluding quasars and bright active galactic nuclei
(AGN) based on their spectral classification by the Sloan Digital Sky Survey
Data Release 7 \citep{SDSS_DR7} in the absolute magnitude range
$-25<M_z<-15$~mag at low redshifts ($0.007<Z<0.27$). We cross-identified the
spectral sample of \emph{SDSS} DR7 galaxies with the UV Galaxy Evolution Explorer
satellite \citep{Martin+05} Release 5 (\emph{GALEX} GR5) catalogue in the
\emph{CASJobs} \citep{Szalay+02} catalogue access systems of \emph{SDSS} and \emph{GALEX}
and rejected the matches separated by more than 3~arcsec on the sky. 

First, we employed the \emph{SDSS} DR7 CasJobs service to select galaxies in the
redshift range from 0.007 to 0.27 from the \emph{SDSS} DR7 spectroscopic sample in
the stripes covered (already or in the survey plan) by the United Kingdom
Infrared Telescope Deep Imaging Sky Survey \citep{Lawrence+07} Large
Area Survey Data Release 8 (\emph{UKIDSS} LAS DR8). We selected only the objects
classified as galaxies by the \emph{SDSS} spectroscopic pipeline (\emph{SpecClass =
2}), that allowed us to reject quasars and prominent broad-line active
galactic nuclei. This list contains 377\,923 sources.

After that we appended both \emph{GALEX} GR5 and \emph{UKIDSS} DR8 data by joining our
reference list of objects with these surveys using the spatial match
criterion, namely the best match within the angular separation of 3~arcsec.
For the \emph{GALEX} data we made use of the pre-calculated cross-match between
\emph{SDSS} and \emph{GALEX} surveys \citep{Budavari+09} accessible through \emph{GALEX} CasJobs
as the \emph{xsdssdr7} table restricting the angular separation to
$<3$~arcsec. For the \emph{UKIDSS} LAS we queried multi-cone search programmatic
access interface with effectively the same parameters for every object from
our initial list. The \emph{SDSS}--\emph{GALEX} join returned 223\,646 galaxies detected
in $NUV$, 144\,639 in $FUV$ ($\lambda_{\mbox{eff}}=155$~nm), and 136\,781 in both filters. The \emph{SDSS}--\emph{UKIDSS}
match contains 176\,868 galaxies detected in the $Y$ band, 178\,806 in $J$,
187\,789 in $H$, 188\,221 in $K$, among them 158\,578 in all four NIR bands.
For 96\,939 of those galaxies we had photometric data from \emph{GALEX} $NUV$,
including 59\,994 with \emph{GALEX} $FUV$ measurements. All the technical
operations on tables were performed with the {\sc stilts} software
\citep{Taylor06}.

We used \emph{SDSS} Petrosian magnitudes (\emph{PetroMag\_*}),
\emph{GALEX} extended source calibrated magnitudes (\emph{nuv\_mag} and
\emph{fuv\_mag}), and \emph{UKIDSS} Petrosian magnitudes
(\emph{PetroMag\_*}) to construct multi-wavelength spectral energy
distributions (SED). Here we notice, that even though SDSS model
magnitudes (\emph{modelMag\_*}) generally have lower formally computed
statistical uncertainties than Petrosian magnitudes, especially in blue
photometric bands, in case of ``blue cloud'' galaxies they are often
hampered by the differences between the observed light distribution and
those assumed (axisymmetrical exponential or de Vaucouleurs) for the
computation of model magnitudes. Petrosian magnitudes may underestimate the
total galaxy flux by 15--20~per~cent in case of face-on de Vaucouleurs
profiles \citep{Yasuda+01}. However, in our case this offset is similar for
\emph{SDSS} and \emph{UKIDSS} data while for the \emph{GALEX} measurements
it is not important because of high photometric uncertainties significantly
exceeding 15~per~cent for red galaxies having the light profile shape
affected by this effect. As long as \emph{GALEX} and \emph{SDSS} contain
photometric measurements in the $AB$ system, but \emph{UKIDSS} magnitudes
are in the Vega system, we applied zero-point transformations available in
the literature \citep{HWLH06} to the NIR magnitudes. We are using integrated
photometry of galaxies, therefore aperture effects have little importance in
the present study and we do not need to apply aperture corrections.

Then, all magnitudes were corrected for the effects of Galactic extinction.
The \emph{UKIDSS} and \emph{SDSS} catalogues provide selective extinction values in all
photometric bands, while for \emph{GALEX} we used the provided $E(B-V)$
value \citep{SFD98} and computed extinctions in UV bands assuming
$A_{NUV}=8.87 \cdot E(B-V), A_{FUV}=8.29 \cdot E(B-V)$.

At this point we created a calibrated photometric sample of low-redshift 
galaxies in 11 bands, from far-UV to NIR. It is easily reproducible at
any workstation with the Internet access, however, due to the data access
policy, the DR4 latest public release of \emph{UKIDSS} catalogues has to be used
instead of DR8.

Systematic uncertainties of \emph{SDSS} point source photometry do not exceed
1~per~cent \citep{Ivezic+04} in $gr$, whereas for extended sources they may
be a few times larger. However, since the red sequence in the optical CMD of
our sample constructed from galaxies populating a large area on the sky is
as tight as 0.03~mag, we conclude that either the systematic errors on
optical magnitudes are within this range, or they are strongly correlated
between $g$ and $r$ bands so that they cannot hamper the results of our
analysis. Statistical uncertainties of \emph{SDSS} photometric measurements are
generally better than 0.015~mag reflecting the spectroscopic target
selection of \emph{SDSS}: galaxies from the spectral sample are at least a few
magnitudes brighter than the limiting magnitude of the photometric survey.
At the same time, the median value of $NUV$ magnitude uncertainties is as
large as 0.15~mag across the whole sample. However, the range of $NUV-r$
galaxy colours is $\sim7.5$~mag compared to $\sim0.9$~mag in $g-r$.
Therefore, the relative ``resolution'' of our analysis per colour range is
very similar in both colours.

\subsection{Computation of $k$-corrections}

As we compare photometric measurements for galaxies at different redshifts,
we have to correct them for the changes of effective rest-frame wavelengths
of filter bandpasses known as $k$-corrections \citep{OS68,HBBE02,BR07}.
Their computation is an important step for obtaining the fully calibrated
homogeneous dataset. Here we provide some details regarding the
$k$-correction computation in \emph{GALEX} UV bands, while the procedure for
optical and NIR filters was exhaustively described earlier \citep{CMZ10}.
The importance of accurate $k$-correction computation is illustrated by the
fact that earlier studies of galaxies in the $(NUV-r, g-r)$ colour--colour
diagram \citep{Yi+05} did not report a sequence of galaxies which would be
obvious if one took a galaxy sample in a narrow redshift range.

Due to high sensitivity of UV fluxes to the recent SF and mass
fractions of young stars as little as 1~per~cent, the UV-to-NIR SED of a galaxy
usually cannot be precisely represented by a single simple stellar
population, that is, a population of stars of the same age and metallicity.
Therefore, $k$-corrections cannot be computed by the single SSP fitting as
it can be done in optical and NIR bands \citep{CMZ10}. Other effects, such as
a non-thermal emission from a moderate-luminosity active galaxy nucleus, or
emission in certain spectral lines, create additional difficulties in the UV
bandpasses. Most of these effects were tackled and successfully taken into
account using the nonnegative matrix factorization \citep{BR07}. In the same
work, the authors computed 5 synthetic template spectra representative of
different galaxies and galaxy components.

Here we used the 2-step process to compute $k$-corrections for our galaxies.
First, we constructed a sub-sample including some 25\,000 galaxies detected
in all 11 bands with high signal-to-noise ratios in the UV bands. We then
fitted their SEDs using a non-negative linear combination of 5
representative templates \citep{BR07} attenuated using the Cardelli extinction
law \citep{CCM89} leaving the colour excess $E(B-V)$ a free parameter.

Second, we used the {\sc topcat}\footnote{\url{http://www.star.bris.ac.uk/~mbt/topcat/}}
table manipulation software \citep{Taylor05} to visualise obtained
$k$-corrections as functions of redshifts and various observed colours, as
we did for optical and NIR bands \citep{CMZ10}. Similarly, we found that UV
$k$-corrections could be precisely approximated by low-order polynomial
functions of redshifts and certain colours. The best filter combinations are
$NUV-g$ and $FUV-u$ for the $NUV$ and $FUV$ bands respectively with standard
deviations of the surface fitting residuals of about 0.08~mag and 0.15~mag.
Then, we used these approximations to compute $k$-corrections for all
galaxies in our sample. The newly obtained approximations of $k$-corrections
are available from the new version of the ``$k$-corrections calculator''
service\footnote{\url{http://kcor.sai.msu.ru/UVtoNIR.html}}.

\section{The colour--colour--magnitude relation and its properties}

We inspected the combined \emph{GALEX}--\emph{SDSS} dataset visually using {\sc topcat} in
three dimensions ($M_z$, $NUV-r$, $g-r$) and detected a very thin continuous
distribution of both, blue and red galaxies along a smooth surface with very
few outliers. Then we approximated it with a low order two-dimensional
polynomial (see Appendix~B for details). Given the range of observed $g-r$
colour of $\sim$0.9~mag, the dispersion of the $g-r$ residuals that decreases
from 0.07 to 0.03~mag going from blue to red $NUV-r$ colours without
significant dependence on the luminosity at $M_z<-17.5$ makes it the
tightest known photometric relation of galaxies.

At lower luminosities, the dispersion of the residuals increases. In our
case this can be explained by significantly lower number of objects due to
the spectroscopic target selection algorithm used in \emph{SDSS} and also by the
poor quality of photometric measurements, because dwarf galaxies have lower
mean surface brightness values than giants and, consequently, their
magnitudes cannot be precisely measured in relatively shallow wide-field
surveys that we used to construct our catalogue. An additional factor
increasing the scatter is the peculiar motions of galaxies inside clusters
and groups not taken into account which hamper the Hubble law distance
estimates.

\begin{figure}
\includegraphics[width=\hsize]{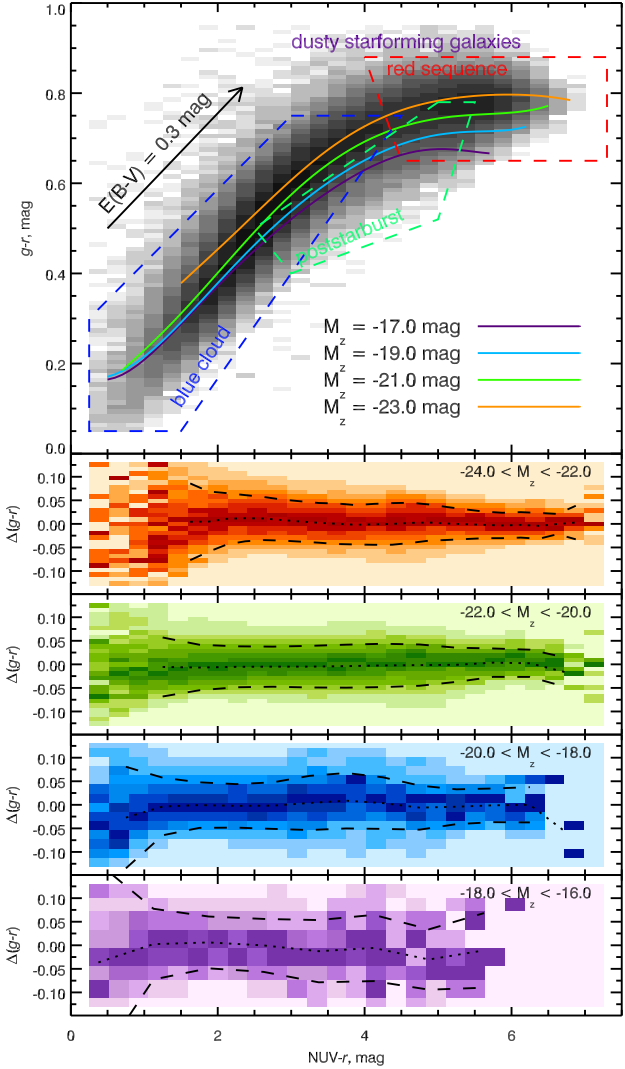}
\caption{The projection of the colour--colour--magnitude relation on-to the
colour--colour plane. The upper panel demonstrates the logarithm of the
number density plot in grayscale with solid lines showing the relations for
galaxies of constant luminosity derived from the best-fitting polynomial
surface equation. Four bottom plots show fitting residuals in different
magnitude ranges as density plots with dashed lines indicating
their $\pm1 \sigma$ levels, which are normalised to the maximum value in
every $NUV-r$ bin. Residuals for the two low-luminosity intervals are
computed with coarser binning compared to the brighter galaxies in order to
account for lower object counts at those luminosities. Red sequence, blue cloud, 
and the loci of certain types of outliers are identified. The direction of 
internal extinction is shown with a vector. \label{fig_colcol}} 
\end{figure}

\begin{figure}
\includegraphics[width=\hsize]{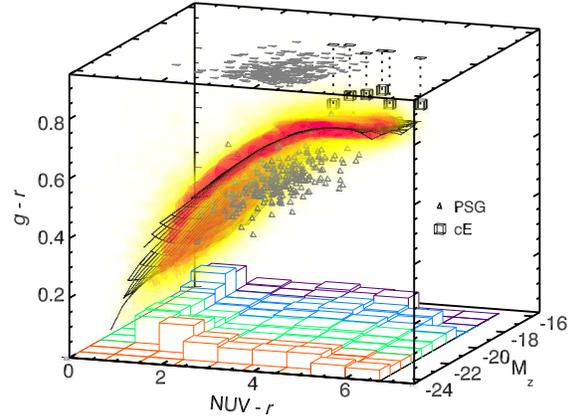} 
\caption{The 3-dimensional distribution of galaxies in the ($NUV-r$,
$M_z$, $g-r$) space. The 3D plot presents
the density distribution of 225\,000 galaxies in the
colour--colour--magnitude space with the increasing density going from
yellow to red, the best-fitting polynomial surface as a mesh grid immersed 
in it, and standard deviations of the fitting residuals shown 
as bars in the ($NUV-r$, $M_z$) plane with their colours corresponding to
the magnitude ranges in Fig.~\ref{fig_colcol}. PSGs \citep{Goto07}, and compact 
elliptical galaxies (cE) are shown as tetrahedra and cubes. The top face 
of the plot demonstrates their projected positions on-to the ($NUV-r$, $M_z$) plane. 
\label{fig_3d}}
\end{figure}

The 3D distribution of galaxies and its projection onto the ($NUV-r$, $g-r$)
plane are shown in Fig.~\ref{fig_colcol},~\ref{fig_3d}. The choice of the
$z$ band for the $M$ axis is not that important: the relation behaves
similarly when using $riz$ or NIR absolute magnitudes. We selected the
\emph{SDSS} $z$ band for presentation purposes as the $z$ luminosity range is
slightly higher than in the $r$ band.

\citet{Yi+05} presented a ($NUV-r$, $g-r$) colour--colour plot for galaxies
and stars in their fig.~1. However, the galaxy photometric measurements were
not properly $k$-corrected as authors did not possess multi-band photometry,
therefore no tight colour relation was revealed.

Rather tight relation in the ($NUV-r$, $g-r$) colour--colour diagram for
normal galaxies was mentioned in \citet{Schiminovich+07}. Then
\citet{Salim+05} and \citet{HGM08} attempted to use both, optical and
near-UV information to study the star formation histories (SFHs) and dust
effects and the effects of environment on the evolution of galaxies.
However, the residual scatter of objects from this relation still remains
high (an order of 0.15~mag in $g-r$) due to the dependence of both galaxy
colours on luminosity. Adding the absolute magnitude as the third dimension
decreases the scatter by a factor of 3 in the absolute magnitude range ($-25
< M_r < -15$~mag). It is remarkable, that the relation in the 3D space is
followed by star-forming as well as passively evolving galaxies.

\subsection{Connection to morphology}

\begin{figure}
\includegraphics[width=\hsize]{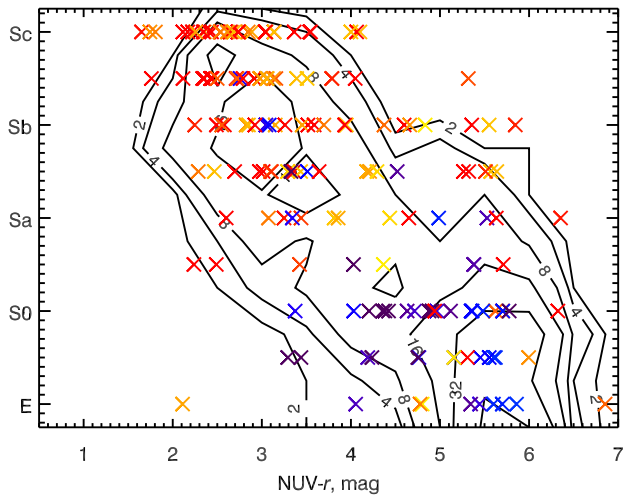}
\caption{Connection between visually determined galaxy morphology
\citep{Fukugita+07} and a $NUV-r$ colour. Black contours correspond to 
galaxies following the relation (numbers are for the counts), 
while individual outliers beyond $1\sigma$ are shown as crosses with the colours
representative of the deviation from the relation in the $g-r$ colour:
yellow to red for galaxies above the relation, blue to violet for objects
below it. \label{fig_morph}}
\end{figure}

To explore the connection of the colour--colour--magnitude relation to
galaxy morphology, we employed the morphological catalogue of \emph{SDSS} galaxies
\citep{Fukugita+07} available through the Vizier
service\footnote{\url{http://vizier.u-strasbg.fr/}}, which contains
morphological types for 1\,465 intermediate luminosity and giant galaxies at
$Z>0.03$ from our sample. We see a continuous change of morphological types
along the surface in the $NUV-r$ direction with a typical dispersion of
0.7$\dots$0.8 Hubble type. At the same time, the optical $g-r$ colour turns
to be a very bad morphological indicator: the red sequence region contains
galaxies of all morphologies from ellipticals to $Sc$ late-type spirals.

The two-dimensional histogram of morphologies vs $NUV-r$ colours is
displayed in Fig.~\ref{fig_morph}. One can see that the S0a and Sa galaxy
types span a very broad range of colours demonstrating the difficulties of
the visual classification of early-type disc galaxies. It
corresponds to a simple linear correlation between the $NUV-r$ and a Hubble
type which can be expressed as:
$$ \mbox{TYPE} = 6.6 - 1.1 \cdot (NUV-r), $$
\noindent where the ``TYPE'' values will correspond to the Hubble types as
$<0$ for \emph{E}, 1 for \emph{S0}, 2 for \emph{Sa}, 3 for \emph{Sb}, 4 for
\emph{Sc}, higher values for \emph{Irr}. Red outliers ($NUV-r \gtrsim 4$~mag)
above the surface (yellow and red points in Fig.~\ref{fig_morph}) mostly
have later types, i.e. spiral galaxies, while outliers below it (blue and
violet points in Fig.~\ref{fig_morph}) have earlier types compared to
galaxies on the sequence with similar $NUV-r$ colours.

The connection between the morphology and the luminosity suggesting that
more luminous galaxies have earlier morphological types is much looser and
may be affected by the selection effects in our sample. 

\subsection{Outliers from the relation}

We identify several classes of outliers from the relation comprising
$\sim$2~per~cent\ of the total sample (see
Fig.~\ref{fig_colcol},~\ref{fig_3d}).

\begin{enumerate}

\item Early-type PSGs selected from the catalogue of H$\delta$-strong
galaxies \citep{Goto07} (359 matches with our sample) populate a region
$\sim$0.15~mag below the surface in $g-r$ spanning $3 \lesssim NUV-r
\lesssim 5$~mag colours explaining the nature of ``blue early-type
outliers'' from the morphology--$(NUV-r)$ relation described above. These
are galaxies with truncated or multi-modal SFH
where the last strong star formation episode has just been finished. The
passively evolving newly formed stellar population reddens much faster in
the $NUV-r$ colour than in the $g-r$ one so that a PSG at first departs from
the blue part of the relation (right in Fig.~\ref{fig_colcol}) and notably
later (after 2.5--3~Gyr) moves up increasing $g-r$ towards the locus
of red sequence galaxies.

\item Compact elliptical galaxies \citep{Chilingarian+09,Price+09} are
residing above the red sequence region of the colour--colour--magnitude
relation at the low luminosity part. A few examples of new cE galaxies are
shown in Fig~\ref{fig_3d}. However, their colours are never redder than
those of the most massive galaxies at the bright red sequence end. This fact
is explained by their formation via severe tidal stripping of more massive
progenitors, most likely early-type disc galaxies, by massive elliptical or
cluster/group dominant galaxies. Progenitors of cEs are stripped in the
innermost regions of galaxy clusters, when the SF is ceased
because their interstellar medium is already removed by the ram pressure
stripping created by the hot intergalactic gas \citep{GG72}. Depending on
the previous SFH, these objects must reside either on the
colour--colour--magnitude relation, or slightly below it, in the PSG locus.
During the relatively fast tidal stripping process lasting about 1~Gyr
\citep{Chilingarian+09}, their stellar population properties and colours
change insignificantly while the mass and, consequently, the luminosity may
decrease by a factor of 10 or more, hence moving a galaxy off the relation
if it was sitting on it. Thus, red cE colours are explained by high stellar
metallicities inherited from their progenitors which is confirmed by
detailed studies of nearby cEs \citep{Rose+05,CB10}. There may be some very
rare intermediate-age cEs originating from PSGs whose colours will be bluer,
however their passive evolution will quickly move them above the
colour--colour--magnitude relation.

\item Dusty starforming galaxies such as edge-on spirals are sometimes found
above the flattened red part of the relation ($NUV-r \gtrsim 4$~mag) also
being consistent with the locus of late-type morphological outliers. Their
positions are explained by the extinction vector direction shown in
Fig.~\ref{fig_colcol}. If internal extinction is very strong then a galaxy
is moved up-right in the diagram and may end up above the locus of red
sequence galaxies.

\item Galaxies with strong ongoing SF but yet small mass fractions of newly
formed stars including ongoing and recent mergers may have very peculiar
colours because of strong nebular emission lines and/or large quantities of
dust.

\item Narrow-line AGNs having low contribution of their nuclei to the total
light in the optical band and hence classified as normal galaxies by the
\emph{SDSS} pipeline may have strong UV excess. We did not apply any particular
filtering to our data to exclude these objects, therefore our sample may be
slightly contaminated by them at a sub-per~cent level.

\item ``Non-physical'' objects: galaxies casually overlapping with either
foreground stars or galaxies at different redshifts create outliers which
may be located in almost any part of the parameter space except the region
very red in ($NUV-r$) and very blue in ($g-r$).

\end{enumerate}

\section{Discussion}

\subsection{Effects of stellar population evolution and internal extinction}

\begin{figure*}
\includegraphics[width=\hsize]{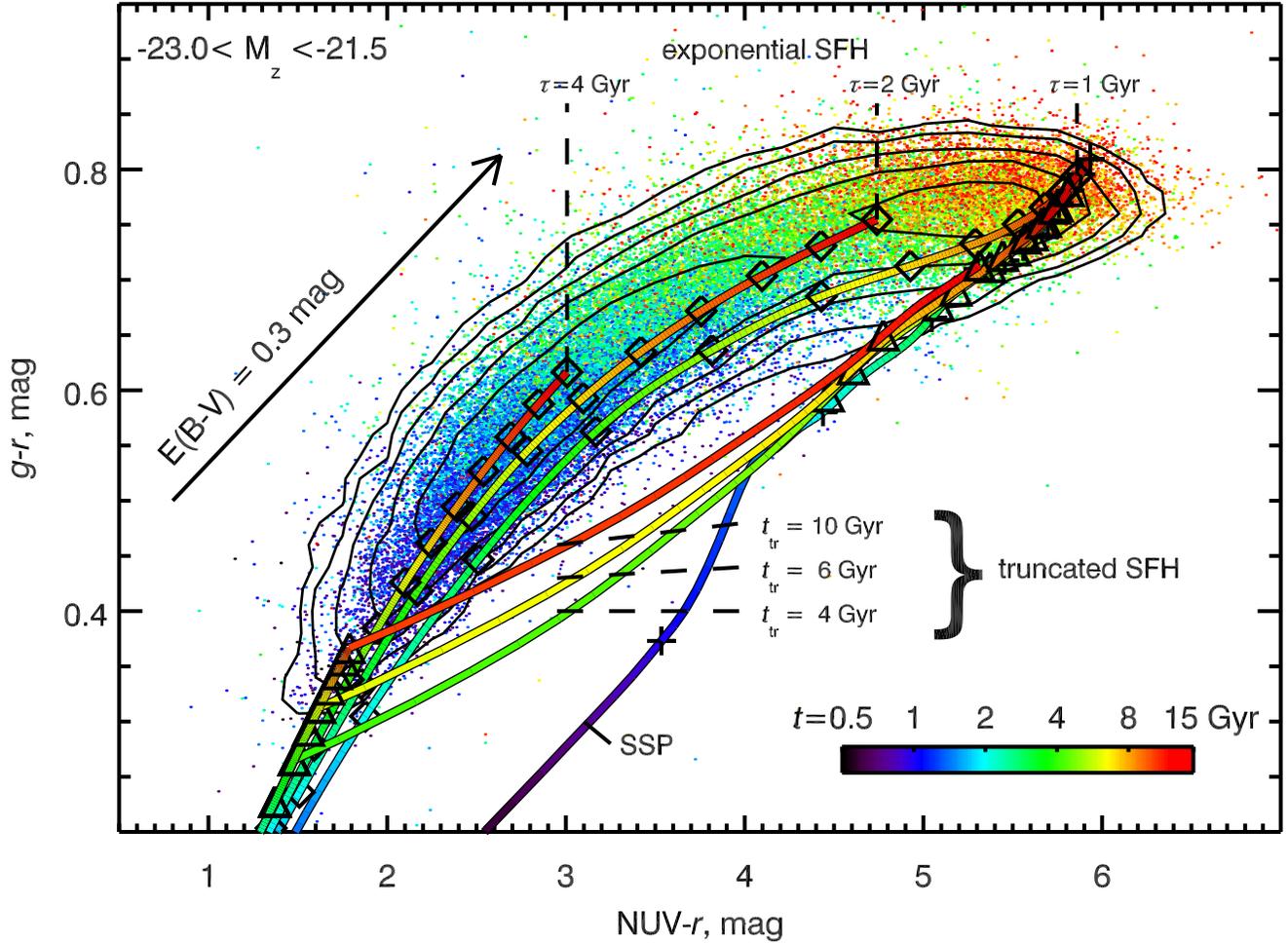} \\
\caption{Connection between galaxy positions on the
colour--colour--magnitude relation and their SSP-equivalent stellar 
populations. The slice of the relation in a narrow luminosity range
($-23<M_z<-21.5$~mag) is displayed in the colour--colour projection. Mean ages
of stellar populations of 40,000 galaxies obtained from the fitting of their
\emph{SDSS} DR7 optical spectra are colour--coded. The evolutionary tracks of 
stellar population models without internal extinction for the solar 
metallicity and various SFH are overplotted (see text). The colours of 
the tracks correspond to the ages of synthetic galaxies formed 12~Gyr ago, 
the ticks on the tracks are given every 1~Gyr. The $NUV-r$ colours in
the models are empirically corrected by $-0.7$~mag (see the text).
\label{fig_colcolage}}
\end{figure*}

The distribution of galaxies in the colour--colour--magnitude space is
governed by three factors: (1) stellar mass, (2) star formation and chemical
enrichment histories including the ongoing SF, and (3) internal extinction.
Therefore, there must be a connection between galaxy positions on the
diagram and their stellar population properties. We fitted a sub-sample of
133\,000 \emph{SDSS} DR7 spectra with simple stellar population (SSP) models using
the {\sc nbursts} technique \citep{CPSA07,CPSK07} and hence obtained their
SSP-equivalent ages and metallicities. The {\sc nbursts} technique includes
the multiplicative polynomial continuum that absorbs flux calibration errors
and makes the fitting insensitive to the internal extinction in a galaxy. We
also notice that \emph{SDSS} spectra obtained in 3~arcsec wide circular apertures
may not be representative of entire galaxies in case of strongly extended
objects with notable gradients of the stellar population properties.
However, we can draw some qualitative conclusions. 

Even in the over-simplified case of SSP-equivalent parameters, we observe a
strong connection between average stellar population properties of galaxies
and their position in the ($M_z, NUV-r, g-r$) parameter space. Galaxies
sitting close to the best-fitting surface exhibit moderate metallicity
gradient as a function of luminosity and almost no variations in the
colour--colour plane except the blue end of the sequence ($NUV-r \lesssim
2.5; g-r \lesssim 0.5$) where the metallicity quickly decreases. The
luminosity--metallicity relation of early-type galaxies known to be
responsible for the tilt of their optical colour--magnitude relation
\citep{KA97} similarly causes the tilt of the colour--colour--magnitude
surface in its red part.

The observed age effects in the colour--colour projection are more
important. At intermediate and high luminosities, the age smoothly increases
along the sequence from $\sim$500~Myr at $NUV-r \approx 1.5, g-r \approx
0.3$ to 13~Gyr at the red end. For low luminosity galaxies, the oldest
SSP-equivalent ages of red galaxies decrease to 10~Gyr at $M_z = -18$~mag
being in accordance with the known anti-correlation of mean stellar
population ages with luminosities of dwarf elliptical galaxies in clusters
\citep{vZBS04,Michielsen+08,Chilingarian+08,Chilingarian09,SLH09,Smith+09}.
Because of the same effect, the upper ``edge'' of the broad red sequence in
the ($NUV-r; M_z$) CMD is strongly tilted at $M_z > -19$~mag.

At all luminosities, there is a notable age gradient across the sequence,
that is, higher $g-r$ colours correspond to older populations. Also, the
dispersion of age estimates ($\log t$) increases while moving towards blue
colours with values totally uncorrelated with colours at the very blue end
of the sequence corresponding to mean stellar ages $t\lesssim500$~Myr.

In Fig.~\ref{fig_colcolage} we present a luminosity slice ($-23.0 < M_z <
-21.5$~mag) of the sample having the median SSP-equivalent metallicity
$[$Fe/H$]=-0.02$~dex ($\langle[$Fe/H$]\rangle=0.13$~dex) and the median
redshift $Z=0.11$ ($\langle Z \rangle=0.03$) corresponding to the light
travel time of $\sim1.4$~Gyr. At this redshift range, the 3~arcsec wide
apertures enclose a large fraction of light from galaxies. The qualitative
behaviour of mean stellar population ages at other luminosities is similar.

We also present the evolutionary tracks for galaxies having various SFH
types. The models were constructed from the {\sc pegase.2} \citep{FR97}
models computed using the synthetic low-resolution BaSeL stellar library
\citep{LCB97} for $NUV-r$ colours, and $g-r$ colours predicted by another
family of stellar population models \citep{Vazdekis+10} computed using a
large Medium-resolution Isaac Newton Telescope library of empirical spectra
(MILES, \citealp{SanchezBlazquez+06}). The combination of the two families
of stellar population models was essential, as \citet{Maraston+09}
demonstrated that the offsets between predicted and observed colours of red
galaxies in the \emph{SDSS} photometric system was due to the nature of synthetic
stellar spectra used to construct stellar population models. The proposed
solution was to use models based on empirical stellar spectra. The MILES
stellar library used to construct models presented here, has the best
coverage of the stellar atmosphere parameters compared to all other existing
published sources except the high-resolution ELODIE library which, however,
has too narrow wavelength coverage making the computation of $g$ and $r$
colours impossible. The $NUV-r$ colours for all models displayed in
Fig.~\ref{fig_colcolage} were empirically corrected by $-0.7$~mag. This
offset is probably of the same nature as that described by
\citet{Maraston+09}, however no models based on empirical stellar spectra
are available in the NUV yet.

The tracks shown in Fig.~\ref{fig_colcolage} were computed as follows.
First, we computed colours and luminosities of simple stellar populations
having $[$Fe/H$]=$0~dex using the {\sc pegase.2} code for the ages of 30,
50, 100~Myr and further till 17~Gyr with a step of 50~Myr. Second, we
computed $g-r$ colours from the MILES-based models and interpolated them to
the same age grid. The resulting SSP track is shown in
Fig.~\ref{fig_colcolage} -- its young part is strongly offset from the
observed distribution of galaxies towards red $NUV-r$ then joining the main
red sequence concentration at ages $t>5$~Gyr. Third, we integrated the
computed SSP luminosities in different photometric bands up-to 12~Gyr using
two families of an SFH: (a) exponentially declining star formation rate
(SFR) with the three characteristic timescales $t_{\mbox{exp}}=1, 2, 4$~Gyr;
and (b) truncated SFHs: constant SFR until a given moment of time from the
galaxy formation epoch ($t_{\mbox{tr}}=4, 6, 10$~Gyr) followed by the
immediate SF cessation and passive evolution afterwards.

Our galaxy evolutionary tracks are simpler than real galaxies because
they do not include the intrinsic metallicity evolution and other processes
such as gas infall from filaments or satellites, mergers, etc., however, we
can use them for some qualitative conclusions:
\begin{enumerate}
\item SSP models at low and intermediate ages ($t<5$~Gyr) to not have any 
corresponding galaxies observed that suggests that (obviously) none of the 
massive galaxies in our sample was formed recently and quickly. However, it
matches quite well the locus of the oldest red sequence galaxies.
\item The main colour sequence can be explained by galaxies formed
immediately after the Big Bang (about 12~Gyr taking into account the median
redshift of our sample) and having various types of a SFH. In
its main part, the slope of the colour--colour relation and the direction of
the internal extinction match each other very well.
\item A galaxy having a constant SFR will end up near the low blue end of the 
sequence and can be moved up--right along it by the internal extinction. A 
family of exponentially declining SFH with different characteristic 
timescales form a curved sequence well corresponding to the observed
relation. The $NUV-r$ colour evolution pace at $3<NUV-r<5$~mag
anticorrelates with the SFR characteristic timescale (see also
\citet{Wyder+07} for a similar plot in the ($NUV-r, u-r$) colour space
without any selection on the luminosity). Models for lower metallicities
well reproduce the colour--colour relation at lower luminosities suggesting
the universality of exponentially declining SFHs. We stress that this
SFH type is not the only one that is able to explain the observed galaxy
distribution in the colour--colour--magnitude space, however, it is the 
simplest model with the smallest number of free parameters compared to 
other alternatives (e.g. multiple starbursts, an exponentially declined law 
with an additional burst).
\item Galaxies having truncated SFHs have a very short transition phase on
their way to the red sequence region lasting about $1$~Gyr after the SF
cessation when their $NUV-r$ colour reddens radically, by $\sim3$~mag while
the $g-r$ change remains about 0.2~mag. The PSG locus below the main colour sequence is
well matched by this transition phase and their rareness is consistent with
a short duration of the transition.
\item Dusty star-forming galaxies above the sequence at red $NUV-r$ colours
are also explained: they are moved up--right from the sequence following the
direction of the extinction vector.
\item The shapes of evolutionary tracks for galaxies with exponentially
declining SFH clearly shows that the evolution of a majority of galaxies
goes along the relation during 6--8~Gyr. \emph{Therefore, we would expect a
weak evolution of the presented colour--colour--magnitude relation shape at
least up-to a redshift $Z\sim0.9$, although the distribution of galaxies on
it will evolve. This suggests it to be a unique search instrument for
distant galaxy clusters using broad band $giJ$ images.}
\end{enumerate}

\subsection{Colour--colour--magnitude relations in other colour pairs}

We found similar photometric relations in other colour pairs, however they
are more strongly affected by observational biases and galaxy evolutionary
phenomena. Colour--colour--magnitude relations involving only optical colours
are very tight because of strong degeneracies between the colours but for
the same reason have very limited astrophysical applications. For example,
in the ($u-r, g-r, M_r$) colour space the $u-r$ is close to $\sim
2\times(g-r)$ at all luminosities, i.e. they are linearly dependent for most
galaxies, hence virtually no information is added by the third dimension.

\subsubsection{Other near-UV--optical colour combinations}

The colour--colour--magnitude relation remains in place when other optical
colour combinations together with the $NUV$ are used provided that there is
enough wavelength lever in the optical colour to distinguish between red and
blue galaxies. That is, colours like $u-r$, $g-i$, $g-z$, $r-z$ but not
$r-i$ and $i-z$. In Appendix~C we provide figures similar to
Fig.~\ref{fig_colcol} constructed for different colour pairs.

Photometric measurements in the $u$ band have relatively poor quality
compared to $g$ and $r$, therefore the residuals of the $(NUV-r, u-r, M_r)$
relation are about four times larger than those of $(NUV-r, g-r, M_r)$. An
additional factor increasing the scatter is an important difference between
the extinction vector direction and the blue slope of the relation in the
($NUV-r, u-r$) plane compared to ($NUV-r, g-r$) increasing the scatter at
$NUV-r < 4$~mag.

The $(NUV-i, g-i, M_i)$ relation has fitting residuals about 50~per~cent
higher than the $(NUV-r, g-r, M_r)$ one, although one would expect them to
be similar given very high quality of $i$ and $r$ band photometry and
similar dependence of these colours on the stellar population evolution. We
explain this by higher uncertainties of the $k$-correction computation in
the $i$ band connected to a broad range of the H$\alpha + [$N{\sc ii}$]$
emission line strength in our galaxies. In our sample, except the very
low-redshift objects ($Z<0.03$), the observed-frame $i$ band may be
contaminated by the H$\alpha + [$N{\sc ii}$]$ emission in a galaxy which may
vary a lot from object to object. However, during the $k$-correction
computation we rely on some average line strength provided by the template
spectra of \citet{BR07}. Therefore, objects with very weak or very strong
emission lines will have their observed $r-i$ colours redder or bluer than
what is predicted by the templates and what was included in the 2D-polynomial
approximations of $k$-corrections. Hence, the $i$ band $k$-correction values
may be biased. Even though, given a $\sim$50~per~cent larger range of $g-i$
colours compared to $g-r$, the relation in this colour space may be used in
the same way as the one presented in Section~3.

The ($NUV-z$, $g-z$) colour combination provides another very good
alternative to $(NUV-r, g-r, M_r)$, however with $\sim$80~per~cent larger
scatter because of lower photometric quality in the $z$ band compared to $r$
in the \emph{SDSS}. The remarkable features of this relation are: (a) a notably
higher luminosity tilt in the red sequence region and (b) low residuals in
the blue part of the relation caused by even better coincidence of the
colour change direction due to the stellar population evolution and internal
extinction than in the ($NUV-r$, $g-r$) plane.

The ($NUV-z$, $r-z$) colour pair starts to suffer from very similar
behaviour of $r$ and $z$ burdened by relatively high photometric errors in
the $z$ band. Therefore, although the $r-z$ colour range is nearly the same
as that in $g-r$, the dispersion of the residuals is notably higher that
complicates the usage of the ($NUV-z$, $r-z$, $M_z$)
colour--colour--magnitude relation.

\subsubsection{Combinations including Far-UV and NIR colours}

\emph{GALEX} $FUV$ measurements have on average much worse quality than the $NUV$
ones because of the lower detector sensitivity and also lower fluxes for
intermediate-age and old galaxies. However, we still can see similar
colour--colour--magnitude relations if we use $FUV$ magnitudes instead of
$NUV$ although with higher residuals especially in the low-luminosity part
of the relation where the mean surface brightness of galaxies decreases. The
computed $k$-corrections also have higher uncertainties in $FUV$ as well as
internal extinction effects introducing additional scatter.

The red sequence region in the colour--colour projection extends from
$\sim$2.5~mag in $NUV-r$ to $\sim$4~mag in $FUV-r$ due to even higher
sensitivity of $FUV$ colours to small fractions of young stars. However, all
combinations involving $FUV$ magnitudes are sensitive to the UV upturn in
old early-type galaxies \citep{Code69,BCO82} likely caused by the stellar
evolution \citep{YDO97}, which results in the ambiguity of the relation in
the red sequence region. That is, after some ``turning point'' (7--8 Gyr),
the $FUV-r$ colour becomes bluer when the stars are getting older.

We used NIR \emph{UKIDSS} photometry to test the existence of photometric relations
in the combinations involving optical-NIR colours. None of the combinations
except ($NUV-Y$, $g-Y$, $M_Y$) provides a relation having similar tightness
to what we detected in the optical colours: the fitting residuals are of an
order of 0.2~mag or larger. 

It is known (see e.g. \citealp{Maraston05}) that $JHK$ colours are
sensitive to AGB stars presenting in intermediate-age stellar populations,
and that at certain ages (1--2~Gyr) the optical--NIR colours ($g-H$ or
$r-H$) are dominated by them being redder than the colours of old stellar
populations by a few tenths of a magnitude. Then, given a much larger range
of e.g. $NUV-H$ than that of $r-H$, this excess will be significant, and it
will strongly depend on the SFH of a given galaxy, so that galaxies with
intermediate $NUV-H$ colours having different SFH families may have
significantly different $g-H$ or $r-H$ colours smearing out the
intermediate-to-red part of the relation except its very red end. In
addition, optical-NIR colours are more sensitive to the metallicity than the
optical ones. Hence, the natural relatively low metallicity spread of
galaxies at a given luminosity will introduce high scatter of their $g-H$ or
$r-H$ colours. Because of the AGB phase, for truncated SFHs, the colour
evolution in the ($NUV-H$, $g-H$) plane will also be more complex than in
the optical colour and it will strongly depend on the truncation time.

\subsection{Redshifts from three photometric points}

This section of the paper aims at an independent mathematical proof of
the existence of the tight UV--optical colour--colour--magnitude relation
of galaxies. The detailed discussion of the technique and its practial
applications to the existing photometric survey data will be provided in a
separate paper.

The mathematical consequence of the relation and smooth dependencies of
$k$-corrections on observed colours is the possibility of the existence of
a univocal functional dependence of a redshift on observed colours and
magnitudes of galaxies. Such a dependence, if found, would confirm the
existence of the universal colour--colour--magnitude relation. Importantly,
it arises from a non-zero curvature of the colour--colour--magnitude surface
and significantly different colour--magnitude distributions for the two
colours used. In a degenerated case, e.g. $(u-r, g-r, M_r)$ where the two
colour--magnitude distributions are very close to the linear dependence and,
therefore, galaxies reside on a surface very similar to a plane, the
photometric redshift determination becomes impossible.

For the following computations we define the two subsamples from the
main galaxy sample extended to the redshift $z=0.52$ (270\,016 objects) by
using their $k$-corrected $g-r$ colours, \emph{red galaxies} ($g-r >
$0.73~mag, 77\,070 objects) and \emph{blue galaxies} ($g-r < $0.7~mag,
167\,157 objects) excluding a small fraction of objects in the ``green
valley''. These samples were again separated into a low- ($z \leq 0.25$) and
high-redshift ($z>0.25$) parts containing 214\,770 (32\,317 \emph{red} $+$
160\,310 \emph{blue}) and 56\,275 (45\,365 \emph{red} $+$ 7\,227
\emph{blue}) galaxies correspondingly. Here, most of blue galaxies in the
high-redshift sample come from the deep \emph{SDSS Stripe 82} imaging
\citep{SDSS_DR4} and the fraction between \emph{blue} and \emph{red}
galaxies clearly demonstrates the target selection algorithm of \emph{SDSS}
biased towards luminous red galaxies at intermediate and high redshifts.

\begin{figure}
\begin{center}
\includegraphics[width=\hsize]{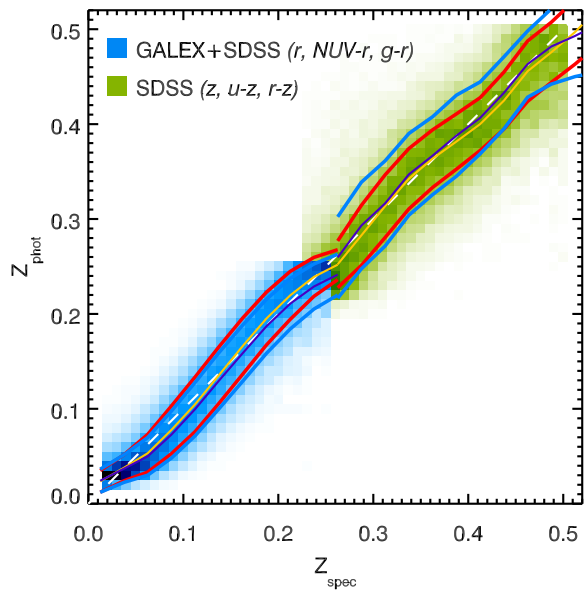}
\end{center}
\caption{Recovery of galaxy distances from three photometric points.
Density histograms of photometric redshift ($Z_{\mbox{phot}}$) determination by fitting
their spectroscopic redshifts ($Z_{\mbox{spec}}$) as a polynomial function
of three parameters: an observed magnitude and two observed colours
corrected for the Galactic extinction. The blue plot displays a sample of
214\,000 low-redshift galaxies ($0.03<Z<0.25$) with NUV, $g$, and $r$
photometric measurements from \emph{GALEX} and \emph{SDSS}. The green plot
is sample of 56\,000 intermediate-redshift galaxies ($0.2<Z<0.52$) whose
redshifts $Z_{\mbox{phot}}$ were determined using $u$, $r$, and $z$ \emph{SDSS} DR7 photometric data.
Red and blue solid lines denote median of $Z_{\mbox{phot}}$ and the
standard deviations of $(Z_{\mbox{phot}}-Z_{\mbox{spec}})/(1+Z_{\mbox{spec}})$ residuals
for \emph{blue} and \emph{red} galaxy subsamples respectively.
\label{figZphot}}
\end{figure}

We approximated the spectroscopic redshifts of galaxies in our
low-redshift subsample as a three-dimensional polynomial
function of observed $r$, $NUV - r$, and $g - r$, and attempted to recover
the photometric redshifts $z_{\mbox{phot}}$ (Fig~\ref{figZphot}).  The
dispersion $\sigma (\Delta (z))$ of the residuals $\Delta (z) =
(z_{\mbox{phot}}-z_{\mbox{spec}})/(1 + z_{\mbox{spec}})$ of 0.025 together
with catastrophic failure rate (defined as fraction of objects with $\Delta
(z) > 3 \sigma (\Delta (z))$) of $\eta = 0.8$~per~cent
 is comparable to the best available photometric redshift
techniques exploiting multi-band FUV-to-NIR photometry, sophisticated
mathematical and statistical algorithms \citep{WS06} and additional
morphological information \citep{WG08}. For \emph{red} and \emph{blue}
galaxies, the residuals and the catastrophic failure rates were:
$\sigma (\Delta (z))_{\rm{red}}=0.021$, $\eta_{\rm{red}} = 2.2$~per~cent and
$\sigma (\Delta
(z))_{\rm{blue}}=0.024$, $\eta_{\rm{blue}} = 0.7$~per~cent.

Consequently, at higher redshifts when the restframe $NUV$ photometric band
shifts to the optical domain, it should be possible to determine photometric
redshifts precisely using $u-r-z$, $g-r-z$, or $g-i-Y$ broadband photometry. 
We tested this hypothesis with our high-redshift subsample by fitting
their redshifts as a function of observed \emph{SDSS} ($z$,
$u-z$, $r-z$) and obtained the residuals having a dispersion $\sigma (\Delta
(z)) = 0.036$ and the rate of catastrophic failures $\eta =
1.1$~per~cent. For \emph{red} and \emph{blue} galaxies, the
residuals and the catastrophic failure rates were: $\sigma (\Delta
(z))_{\rm{red}}=0.034$, $\eta_{\rm{red}} = 1.0$~per~cent and $\sigma (\Delta
(z))_{\rm{blue}}=0.047$, $\eta_{\rm{blue}} = 1.2$~per~cent. These relatively
large errors are mostly due to the very poor quality of $u$-band Petrosian
magnitudes having typical uncertainties of an order of 0.3~mag for
high-redshift galaxies. We notice here, that if one uses model magnitudes
instead of Petrosian ones, the relation becomes much tighter for the
\emph{red} subsample ($\sigma (\Delta (z))_{\rm{red}}=0.027$), however it nearly
disappears for \emph{blue} galaxies whose $u$-band model magnitudes do not
correspond to their real photometric properties because of light
distributions being very far from regular exponential or de Vaucouleurs
profiles.

The demonstrated possibility of the precise photometric redshift
computation for both, red and blue galaxies with a small fraction of
outliers from three photometric points involving a near-UV and optical
colours proves the existence of the tight unversal colour--colour--magnitude
relation for normal galaxies of all types, not only red sequence objects.

We compare these metrics to the existing photometric redshifts techniques.
Using the ``{\sc le phare}'' photometric redshift code combined with a
template optimisation procedure and the application of a Bayesian approach,
based on the sample of galaxies with 9 individual photometric measurements
of a quality similar to ours, \citet{Ilbert+06} find the dispersion $\sigma
(\Delta (z))$ to be 0.025 and $\eta = 1.9$~per~cent (though we stress that
the authors defined the catastrophic failure limit at the fixed level of
0.15 which corresponded to $6 \sigma$ in their statistics). 
\citet{Mobasher+07} analyse the performance of different photometric
redshift codes on a dataset that comprises 16 photometric points for every
SED in question.  The best result achieved in their study with own method
reaches the dispersion of residuals as low as $\sigma (\Delta (z)) = 0.033$
and $\eta = 2.2$~per~cent. The \emph{SDSS} database provides several
photometric redshift estimates obtained as described in \citet{SDSS_DR5}. 
We have extracted three of them: (1) those obtained from the comparison of
the observed colors of galaxies to a semi-empirical reference set (hereafter
photoz) from the {\it photoz} table, and (2) neural network estimators
derived from galaxy magnitudes (``D1'') and (3) colours (``CC2'') from the
{\it photoz2} table. In general, they perform quite well for {\it red}
galaxies in our high-redshift subsample ($\sigma (\Delta (z))_{\rm red,
photoz} = 0.020$ and $\eta_{\rm red, photoz} = 1.7$~per~cent, $\sigma
(\Delta (z))_{\rm red, D1} = 0.020$ and $\eta_{\rm red, D1} = 1.6$~per~cent,
$\sigma (\Delta (z))_{\rm red, CC2} = 0.023$ and $\eta_{\rm red, CC2} =
1.7$~per~cent). However, similarly to our approach, the quality is worse for
{\it blue} galaxies: $\sigma (\Delta (z))_{\rm blue, photoz} = 0.042$ and
$\eta_{\rm blue, photoz} = 3.4$~per~cent, $\sigma (\Delta (z))_{\rm blue,
D1} = 0.039$ and $\eta_{\rm blue, D1} = 2.0$~per~cent, $\sigma (\Delta
(z))_{\rm blue, CC2} = 0.044$ and $\eta_{\rm CC2} = 1.7$~per~cent. One has
to keep in mind that (1) these techinques use a lot of additional
information (e.g.  morphology, size, etc.) and (2) all the galaxies in the
spectroscopic \emph{SDSS} sample in fact constitute a training sample of
these methods, so one has to check those numbers against massive third-party
redshift surveys.

We also note that since the redshift determination by three
photometric points is a mere mathematical consequence from the
colour-colour-magnitude relation, photometric redshift outliers are
themselves objects that fall aside from the relation, namely PSGs, dusty
starbursts, and AGN.

To summarise, compared to existing photometric redshift techniques,
presented method requires a factor of 3 to 5 more modest investment in
observing time (due to the fact that individual galaxy measurements do not
need to be made in as many photometric bands), being able to provide
redshifts for large samples of galaxies at the same or better level of
accuracy.  Moreover, proposed polynomial evaluation is significantly simpler
from the methodological point of view than $\chi^2$ minimization with
Bayesian priors used in mainstream photometric redshift codes.

There are two main disadvantages of our approach: (1) it is not precise for
non-typical galaxies, i.e. outliers from the colour--colour--magnitude
relation; (2) it works only for those regions of the parameter space, that
are well sampled with spectral redshift measurements. Latter means e.~g.
that it is possible to go beyond the \emph{SDSS} spectral sample magnitude limit
retaining the declared precision of our method if one uses an external
source of spectral redshifts to calibrate the functional relation to work
with the \emph{SDSS} photometry at fainter magnitudes. But without such a
calibration, photometric redshifts estimates for faint galaxies will be
wrong.

Conceptually, the presented multi-dimensional polynomial fit resembles the
training of artificial neural networks sometimes used for the photometric
redshift determination (e.g. \citealp{DAbrusco+07}), though the underlying
machinery is different. In both cases, there is a non-linear transformation
(a 3D-polynomial function in our case or consequent multi-level sigmoid
transformations in case of neural networks) of some input measurements into
the output redshift estimate. And the coefficients of a transformation are
tuned (``trained'') in a way to work as good as possible for the reference
(``training'') dataset. Hence, both methods unlike $\chi^2$ template fitting
family are insensitive to systematic errors of the input data. That is,
the ``templates'' are constructed from the dataset itself and are tolerant
to its problems by construction.

\section{Summary}

We presented the universal very tight colour--colour--magnitude relation in
optical and near-UV filters followed by the vast majority of non-active
galaxies of all morphological types covering at least 8 magnitudes in
luminosity from the sample including 225\,000 low-redshift ($Z<0.27$)
galaxies observed by \emph{SDSS} and \emph{GALEX} surveys. A special case is the
connection of the optical $g-r$ colour to the $NUV-r$ colour and $M_r$
luminosity which we approximated by a low-order polynomial surface with the
residuals of $\sim$0.03--0.07~mag in the entire covered luminosity ($-23.5 <
M_r < -15.5$~mag) and colour ($0 < NUV-r < 7.5$~mag) ranges.

We have demonstrated that there is a strong correlation between the $NUV-r$
colour and the galaxy morphology while for the optical $g-r$ colour this
correlation is much weaker. We identified several classes of the outliers
constituting about 2~per~cent of the total galaxy population and explained
their nature, the most important being rare compact elliptical and
poststarburst E+A galaxies. 

We have performed stellar population modelling and shown that the relation
can be explained by galaxies having constant and exponentially declining
SFHs, while truncated SFHs with very rapid colour evolution after the SF
cessation explain the properties of E+A outliers. The most important
conclusion is the predicted weak dependence of the colour--colour--magnitude
relation shape on a redshift up-to $Z\sim0.9$ that will allow to use it for
search of distant galaxy clusters in 3-colour broad band images.

The existence of such a tight photometric relation suggests the possibility
of the high-precision photometric redshift estimates using only three
photometric points that we have also illustrated. Our empirical photometric
redshift technique using a minimalistic input dataset has a quality
comparable to or better than most published $Z_{\mbox{phot}}$ techniques,
while it is much simpler to implement.

The detailed astrophysical interpretation of the described photometric
relation still has to be done. However, it can be used already as a powerful
galaxy classification and selection instrument based only on their
integrated photometry as well as a tool to search for representatives of
rare galaxy types in photometric galaxy samples.

\section*{Acknowledgments} We acknowledge the ADASS conference series
(\url{http://www.adass.org/}), because this result emerged while preparing
the tutorial for the ADASS-{\sc xx} meeting.  This work would have been
impossible without technologies developed by the International Virtual
Observatory Alliance (\url{http://www.ivoa.net/}) and tools and services
maintained by the participating national VO projects: Euro-VO, VAO,
Astrogrid.  This research has made use of TOPCAT, developed by Mark Taylor
at the University of Bristol; Aladin developed by the Centre de Donn\'ees
Astronomiques de Strasbourg (CDS); the ``exploresdss'' script by G.~Mamon
(IAP); the VizieR catalogue access tool (CDS).  Funding for the \emph{SDSS} and
\emph{SDSS}-II has been provided by the Alfred P.  Sloan Foundation, the
Participating Institutions, the National Science Foundation, the U.S. 
Department of Energy, the National Aeronautics and Space Administration, the
Japanese Monbukagakusho, the Max Planck Society, and the Higher Education
Funding Council for England.  The \emph{SDSS} Web Site is
\url{http://www.sdss.org/}. \emph{GALEX} (Galaxy Evolution Explorer) is a
NASA Small Explorer, launched in April 2003.We gratefully acknowledge NASA's
support for construction, operation, and science analysis for the
\emph{GALEX} mission, developed in cooperation with the Centre National
d'Etudes Spatiales of France and the Korean Ministry of Science and
Technology. Authors acknowledge the Oversun-Scalaxy
(\url{http://www.scalaxy.ru/}) cloud computing provider for resources used to
perform a part of this study. Special thanks to F.~Combes, A.~Graham, and
R.~Ibata for useful discussions and suggestions.

\bibliographystyle{mn2e}
\bibliography{UV_optical_CMD}

\appendix

\section{Validation of the result}

Two factors may in principle lead to the spurious creation of the
colour--colour--magnitude relation presented in this work: (a) sample
selection biased toward specific colours corresponding to the described
surface, (b) serious faults in the $k$-correction computation artificially
bringing most galaxies to that relation.

As far as we apply no selection to the \emph{SDSS} DR7 spectroscopic sample of
galaxies based on their morphology, colours, sizes etc., and \emph{GALEX} is a
full-sky survey providing photometric information for all detected objects,
the only source of selection effects may be the spectroscopic target
selection of \emph{SDSS}. The corresponding algorithms are exhaustively
described \citep{Eisenstein+01,Strauss+02}, and we find no evidences for them
to introduce any biases on galaxy selection that may lead to the creation of
``empty'' regions in the colour--colour--magnitude parameter space. This is
also well illustrated by the broad and very well filled colour distribution
of \emph{SDSS} galaxies \citep{Strateva+01,Blanton+03b,BGB04} and studies of galaxy
luminosity functions based on \emph{SDSS} \citep{Blanton+03a}.

In order to test the existence of potential problems in the $k$-correction
computations, we have conducted two specific tests. For this, we selected
two sub-samples of galaxies in narrow redshift ranges, $\Delta z_1$:
$0.03<z_1<0.05$ (24\,319 galaxies) and $\Delta z_2$: $0.08<z_2<0.10$
(37\,303 galaxies). We chose relatively low-redshift samples because the
\emph{SDSS} targeting algorithm has a limiting magnitude $r=17.77$~mag in a
3~arcsec aperture so that galaxies at higher redshifts do not sample well
the luminosity axis of the parameter space.

\begin{figure}
\includegraphics[width=\hsize]{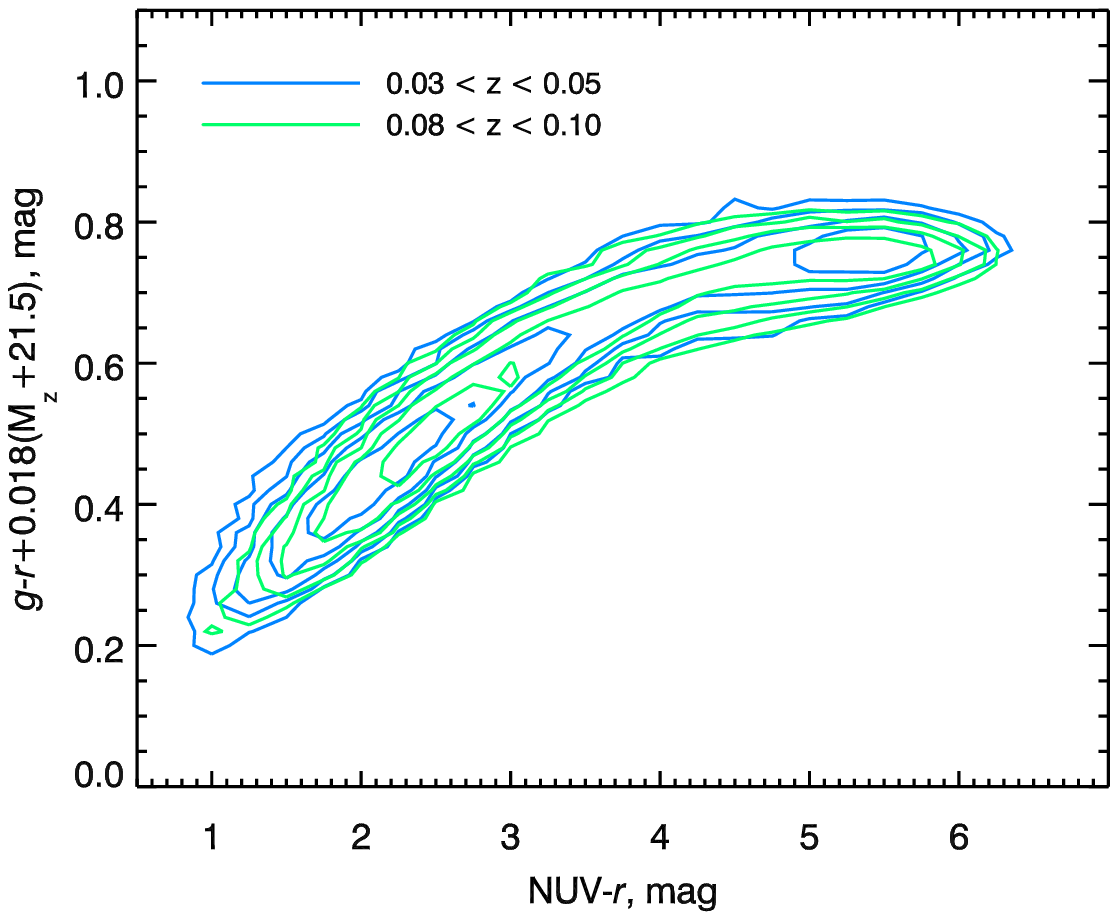}\\
\includegraphics[width=\hsize]{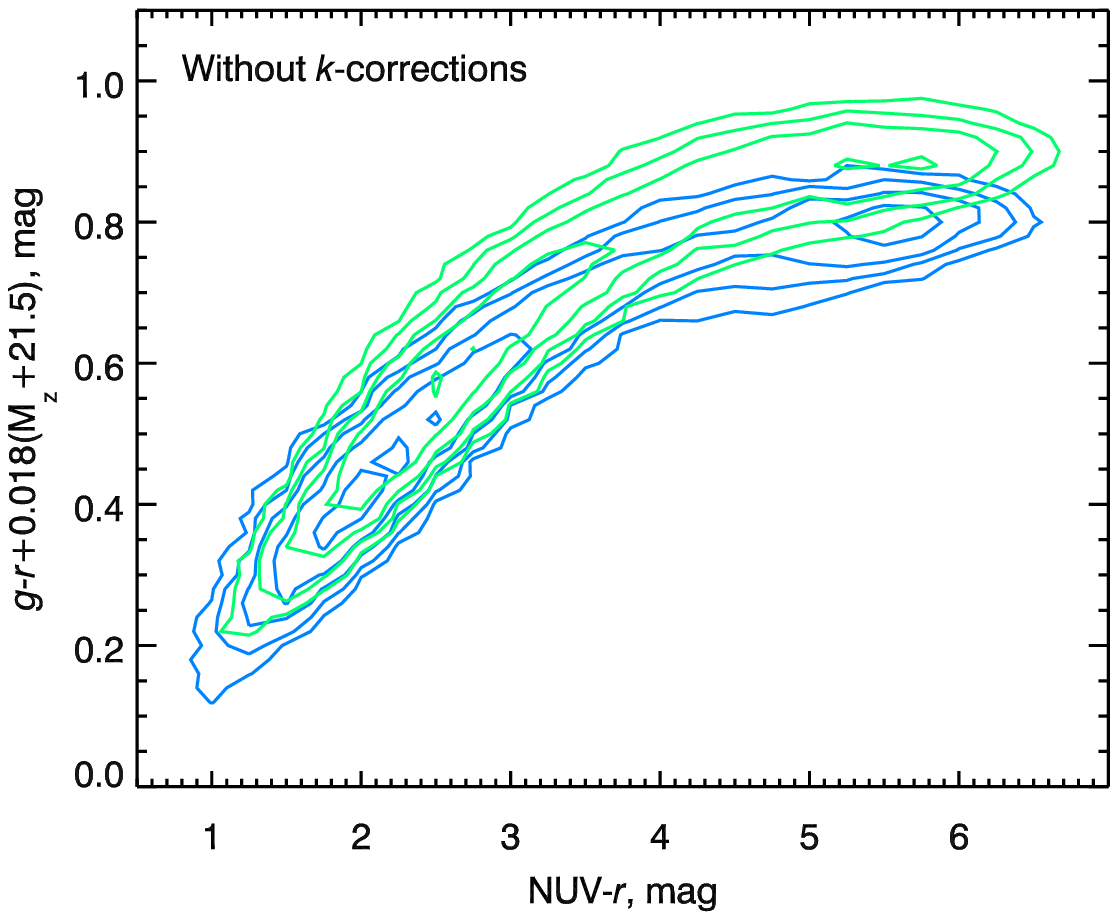}
\caption{Tests with the two sub-samples of galaxies in narrow redshift
ranges. The top panel demonstrates the number density plots for galaxies
in two narrow redshift ranges in the projection shown in green and blue, 
corresponding to the edge-on view of the colour--colour--magnitude relation.
The bottom panel is the same as the top one, but without applying
$k$-corrections to the measurements.
\label{fig2zr}}
\end{figure}

The first test was to compare the colour--colour--magnitude distributions of
galaxies in the two sub-samples. In case of $k$-correction computation
problems, one would expect systematic differences between the two
sub-samples, which we have not detected (see Fig.~\ref{fig2zr}a for an
``edge-on'' view of the relation).

The second test was to entirely disable the $k$-correction computation for
these two sub-samples. Because $k$-corrections do not change significantly
within the narrow redshift sub-samples, but do change between the sub-samples,
one would expect to get two tight colour--colour--magnitude sequences
qualitatively resembling the relation for $k$-corrected magnitudes, but with
some quantitative differences such as the blue cloud slopes and red
sequence positions. We obtained the result exactly as predicted by this
intuitive assumption. The edge-on views of the non $k$-corrected relations
are presented in Fig.~\ref{fig2zr}b.

However, the strongest argument supporting the existence of a universal
colour--colour--magnitude relation in UV--optical colours is the possibility
of computing photometric redshifts using only three observed colours as
demonstrated in the Discussion section.

\section{Fitting surfaces into strongly non-uniform three-dimensional
scattered datasets}

A surface fitting procedure is an essential mathematical component required
to obtain results presented in the paper. The observational photometric
datasets for galaxies from wide-field survey have strongly non-uniform
distribution in the colour--colour--magnitude space due to the superposition
of the complex distribution of galaxies connected to their physics and
various selection effects and observational biases.

Using visual inspection, we revealed a distribution of points in the
colour--colour--magnitude space close to a smooth surface, but applying
standard $\chi^2$-based linear surface fitting techniques did not yield the
results of a reasonable quality because: (a) the density of points in the
NUV-colour--magnitude plane varies by several orders of magnitude while the
individual measurements have comparable quality; (b)
distribution of points around the surface is sometimes significantly
asymmetric and non-Gaussian. The former property of the distribution leads
to the fact that the scarcely populated regions can deviate significantly
from the surface without notable change of the goodness-of-fit as the
best-fitting surface tries to minimize the deviation in the densest regions
of the parameter space. The latter property leads to the biased fitting
results as the $\chi^2$ technique assumes the Gaussian distribution. The
$\kappa$-sigma clipping technique will not solve the problem here because
we are dealing with a large number of points deviating from the symmetric
distribution and not with individual outliers.

To tackle these issues in a simple way, we decided to use a two--step
technique for the surface fitting. First, we defined a fine grid (cell size
of about 0.25$\times$0.25~mag) in the NUV-colour--magnitude plane and in
every bin computed median values of the optical colour being fitted. This
allowed us to pick up the maxima of the (e.g. $g-r$) colour distributions in
every bin and not the mean values, that was critical in order to account for
the asymmetrical distribution of points around the surface. Then we filtered
out the values where the 2D-histogram counts in a given bin were below some
threshold (usually, 5 or 7 galaxies). At the second step, we fitted a
low-order polynomial surface into these median values using a standard
routine fitting linearly the polynomial coefficients and assigning equal
weights to all points remained after the filtering at the first stage. This
way we took into account the strongly non-uniform distribution of galaxies
in the NUV-colour--magnitude plane.

This two--step approach resulted in almost flat distribution of residuals
displayed in Fig.~\ref{fig_colcol} and in the next Appendix.

\section{Colour--colour--magnitude relations in different colour
combinations}

\begin{figure}
\includegraphics[width=\hsize]{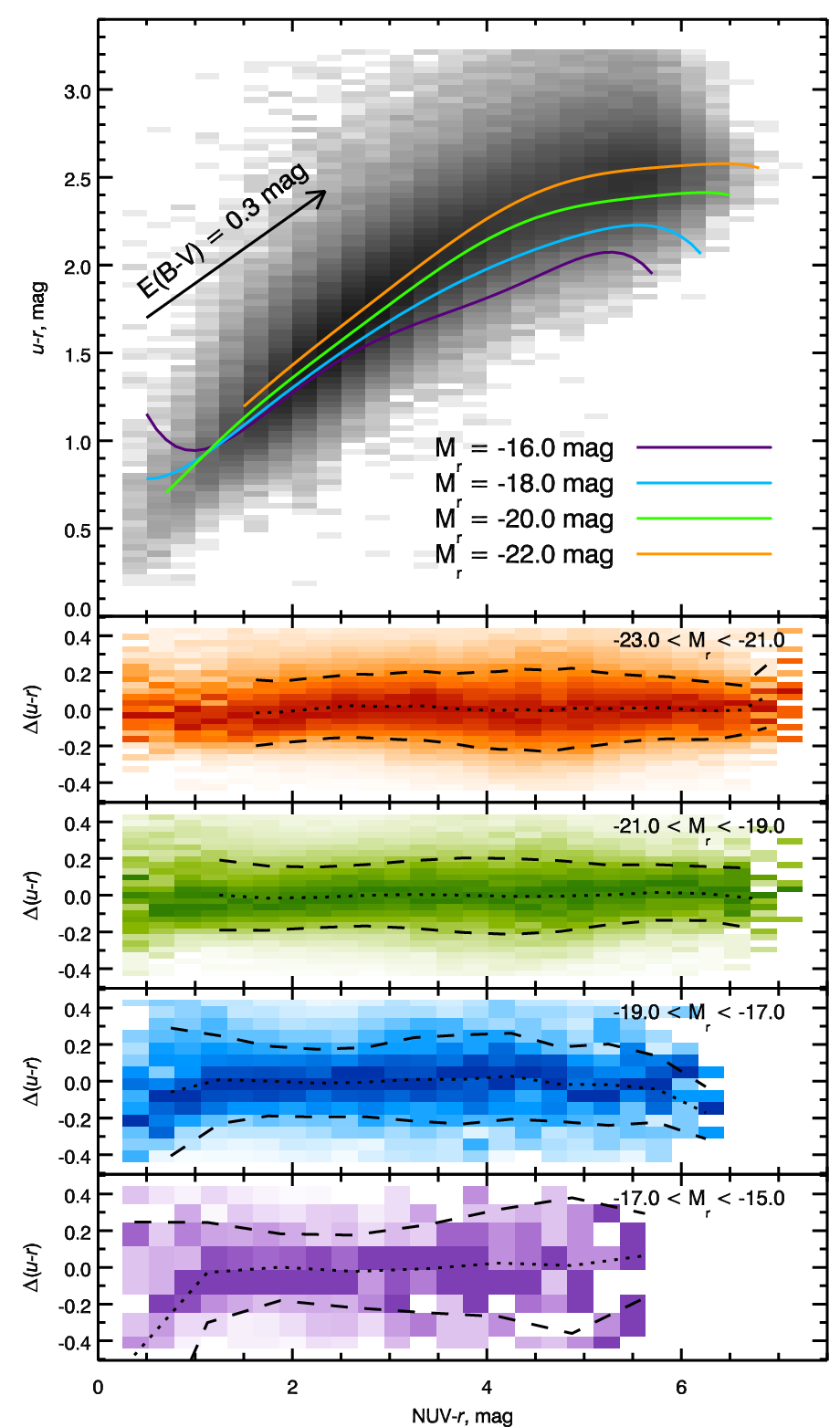}
\caption{The same as Fig.~\ref{fig_colcol} but for the ($NUV-r$, $u-r$, $M_r$)
space.\label{fig_colcol_urr}}
\end{figure}

\begin{figure}
\includegraphics[width=\hsize]{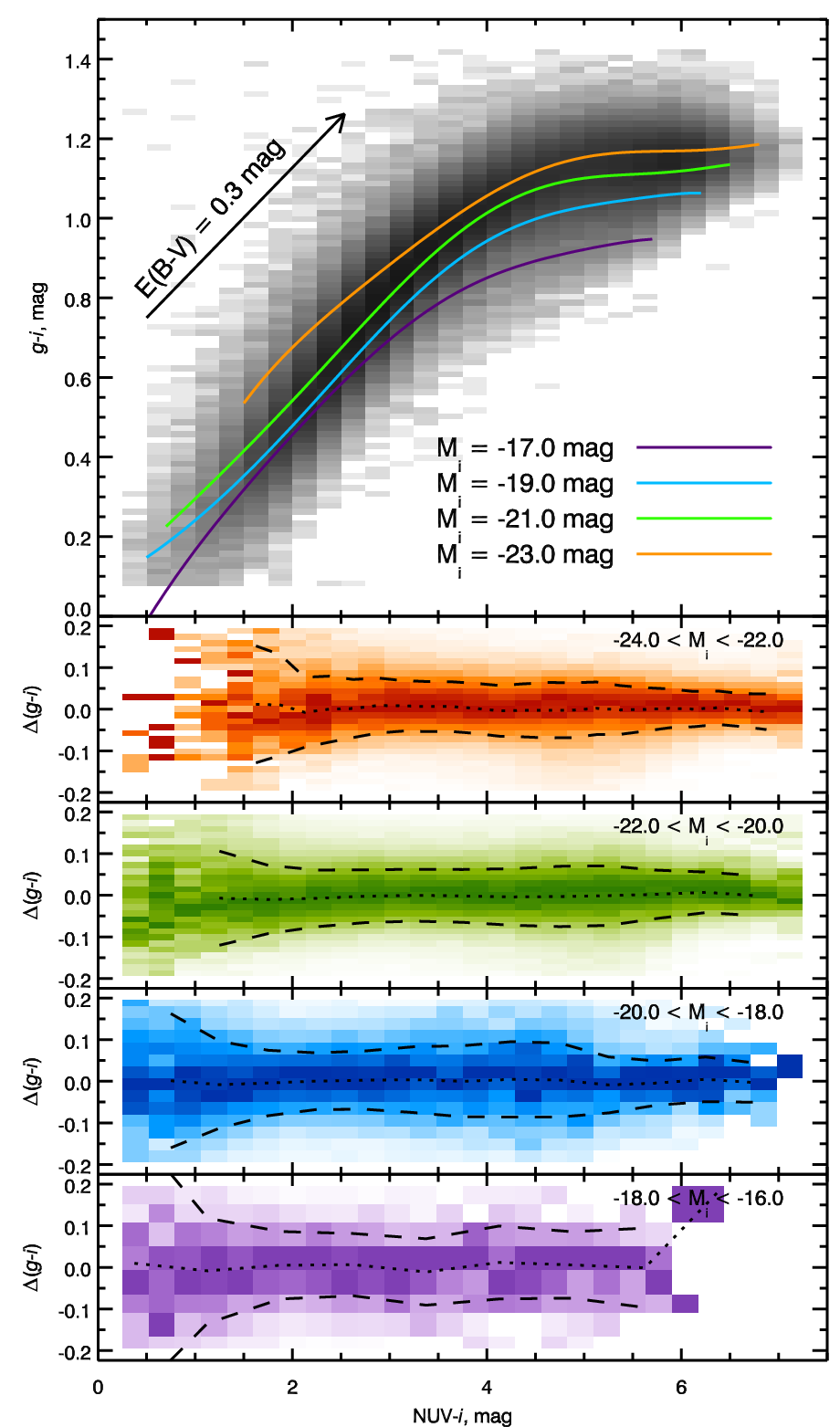}
\caption{The same as Fig.~\ref{fig_colcol} but for the ($NUV-i$, $g-i$, $M_i$)
space.\label{fig_colcol_gii}}
\end{figure}

\begin{figure}
\includegraphics[width=\hsize]{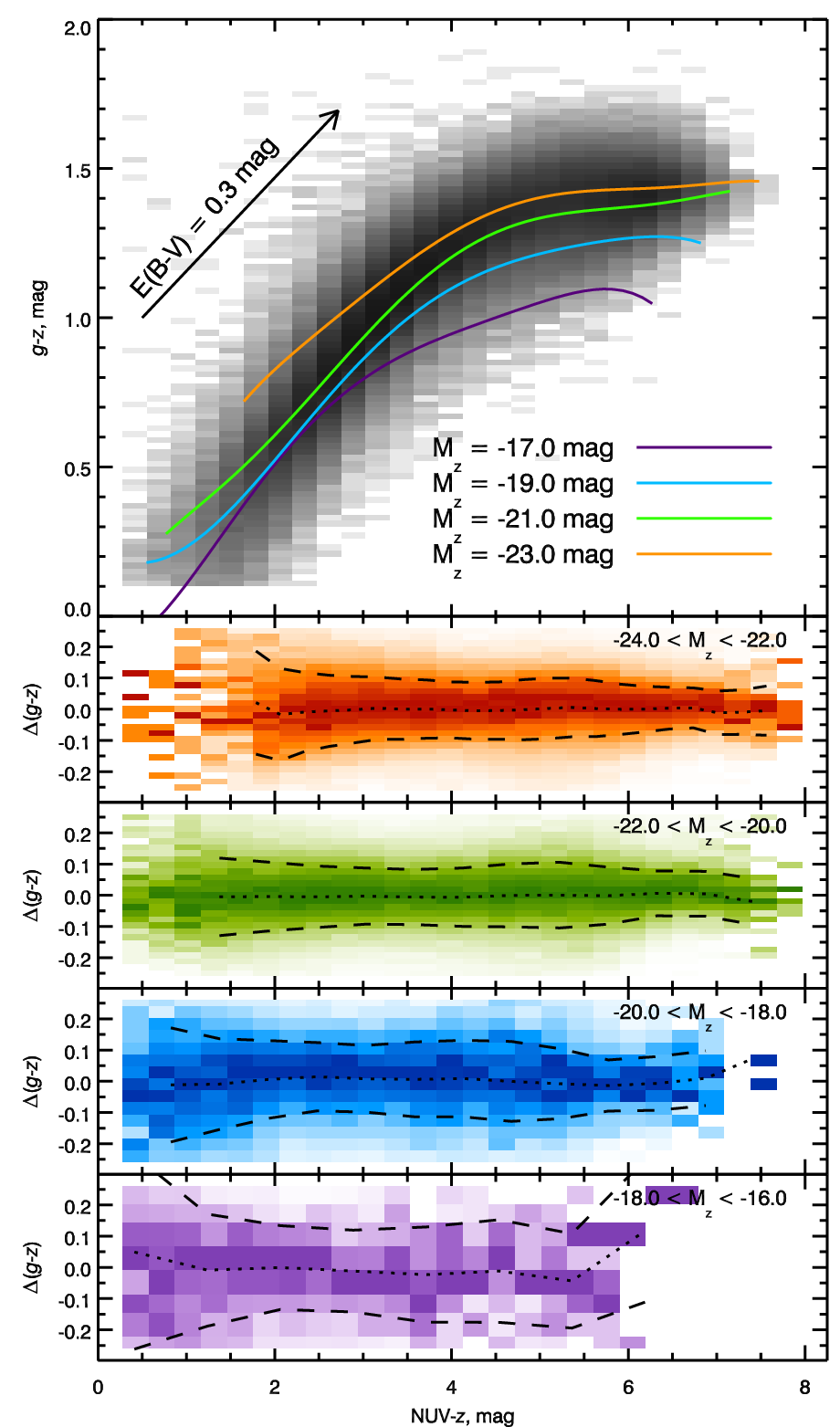}
\caption{The same as Fig.~\ref{fig_colcol} but for the ($NUV-g$, $g-z$, $M_z$)
space.\label{fig_colcol_gzz}}
\end{figure}

\begin{figure}
\includegraphics[width=\hsize]{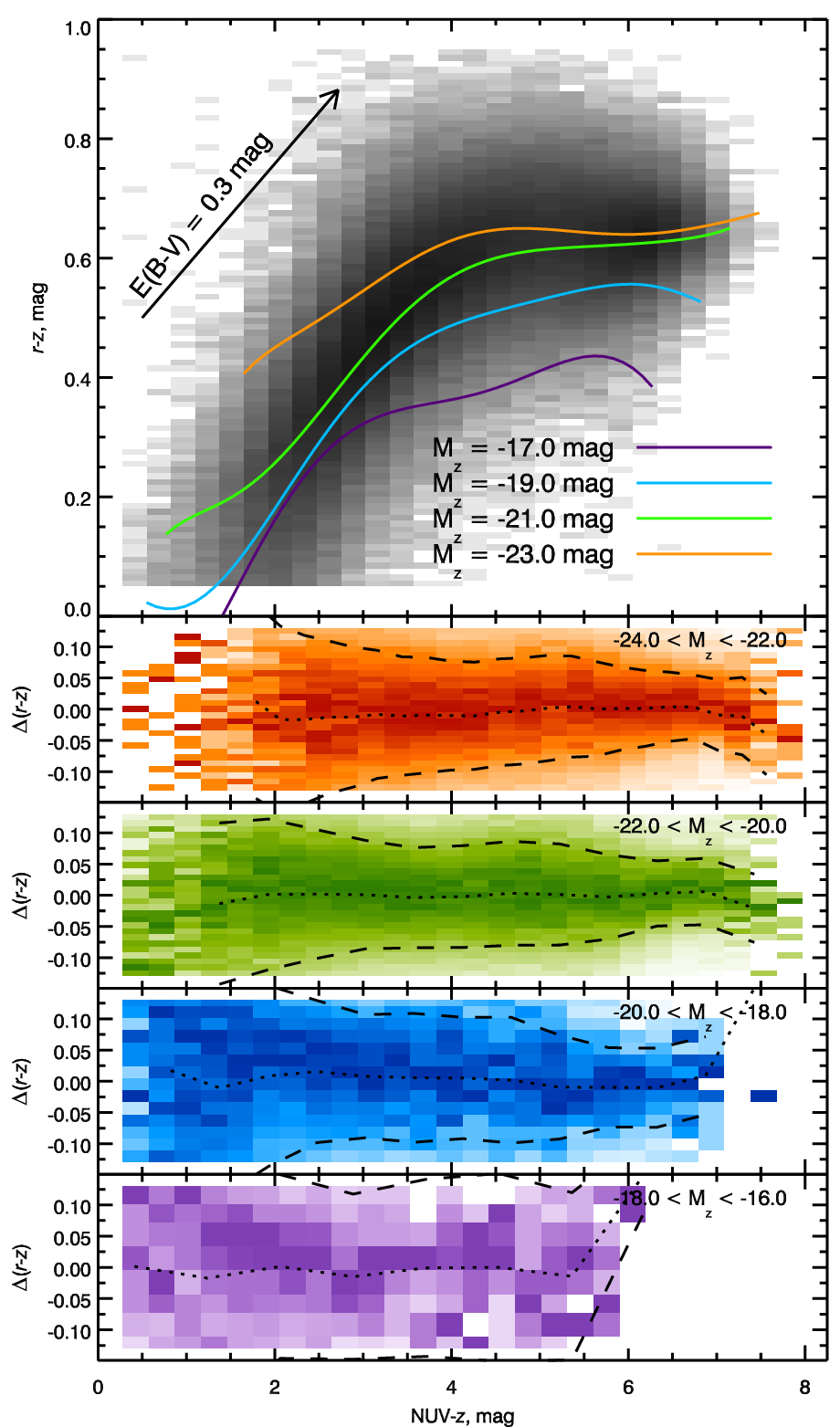}
\caption{The same as Fig.~\ref{fig_colcol} but for the ($NUV-z$, $r-z$, $M_z$)
space.\label{fig_colcol_rzz}}
\end{figure}

In Fig.~\ref{fig_colcol_urr}--\ref{fig_colcol_rzz} we show the
colour--colour projections of the colour--colour--magnitude relation in
different near-UV--optical colours described in Section~4.2. All the fitting
results including the coefficients of the best-fitting surfaces, fitting
residuals in colour--magnitude bins and other essential information for the
usage of the relations are provided in the electronic
form\footnote{http://specphot.sai.msu.ru/galaxies/}.

\label{lastpage}

\end{document}